# Performance and Interpretability of Convolutional, Transformer, and Hybrid Deep Learning Models in Colorectal Histology Classification

Reza Bozorgpour *

College of Engineering & Applied Science, Department of Biomedical Engineering, University of Wisconsin-Milwaukee, Milwaukee, WI, USA

*Corresponding author
E-mail: bozorgp2@uwm.edu

## Abstract

Deep learning has become an important tool in computational pathology, enabling automated analysis of histopathological images. While convolutional neural networks (CNNs) have traditionally dominated this field, transformer-based and hybrid architectures have recently demonstrated promising performance. However, comprehensive comparisons of these approaches for colorectal histopathology remain limited. This study evaluated twelve ImageNet-pretrained CNN, transformer, and hybrid architectures using the Kather colorectal histopathology dataset containing 5,000 image tiles from eight tissue classes. All models were trained using a standardized transfer-learning and fine-tuning protocol and assessed using multiple performance metrics, including accuracy, precision, sensitivity, specificity, F1-score, ROC-AUC, Cohen's kappa, and Matthews correlation coefficient. All evaluated models achieved high classification performance, with accuracies ranging from 93.2% to 97.1%. EVA-02 achieved the highest overall performance (97.1% accuracy, 97.0% F1-score), closely followed by ViT-B/16. Among CNNs, ResNet34 and ConvNeXt-Tiny demonstrated highly competitive performance, achieving accuracies of 96.4% and 96.3%, respectively. Transformer architectures generally produced the strongest results across evaluation metrics, although the performance gap between the best transformer and CNN models was relatively small. Per-class analysis showed consistently strong classification performance across all tissue categories, with Complex Stroma representing the most challenging class. Overall, transformer-based architectures achieved the highest predictive performance, whereas modern CNNs provided a favorable balance between accuracy and model complexity. These findings provide a comprehensive benchmark of major deep learning paradigms for colorectal histopathology classification.

## Introduction

Colorectal cancer (CRC) is one of the most prevalent malignancies worldwide and remains a major cause of cancer-related morbidity and mortality. It ranks as the second most common cancer among women and the third among men globally [1]. Although incidence rates vary according to geographic region, age, and sex, the global burden of CRC is expected to increase substantially, with projections indicating an approximately 80% rise in new cases by 2035 [2]. This trend has been linked largely to lifestyle and environmental factors, particularly dietary patterns associated with increased cancer risk [3]. While most CRC cases occur sporadically, hereditary factors contribute significantly to disease susceptibility in a substantial proportion of patients [4]. Importantly, CRC represents a heterogeneous disease rather than a single pathological entity, encompassing tumors with diverse histological, molecular, and clinical characteristics that influence disease progression, treatment response, and patient outcomes.

The growing incidence of CRC has placed increasing demands on pathology services, where histopathological evaluation remains the cornerstone of diagnosis [5, 6]. Despite major advances in molecular diagnostics, microscopic examination of tissue sections stained with Hematoxylin and Eosin (H&E) continues to serve as the primary method for identifying colorectal lesions and assessing their pathological features [7]. Additional techniques, including immunohistochemistry (IHC), in situ hybridization (ISH), and molecular assays, are frequently employed to provide complementary diagnostic and prognostic information [8]. At the same time, widespread screening programs have increased the detection of precursor lesions and early-stage abnormalities, expanding the range of tissue patterns encountered in routine practice and creating new diagnostic challenges, particularly when distinguishing between premalignant and malignant lesions [9]. Combined with the growing requirement to evaluate an increasing number of morphological and molecular biomarkers, these developments have contributed to a more complex and time-intensive diagnostic workflow [10].

Another challenge arises from the inherently interpretive nature of histopathological assessment. Diagnostic decisions rely heavily on the recognition of visual patterns within tissue architecture and cellular morphology, a process that is strongly influenced by individual expertise and experience [11, 12]. Although highly trained pathologists achieve remarkable diagnostic accuracy, variations in interpretation may occur between observers or even during repeated assessments by the same observer. Such variability has motivated the search for computational approaches capable of supporting more objective, reproducible, and efficient diagnostic decision-making [9, 13].

The emergence of digital pathology has created the technological foundation for these advances. The routine conversion of glass slides into high-resolution whole-slide images (WSIs) [14-17] has enabled the large-scale digitization of histopathological specimens, generating extensive datasets suitable for computational analysis [18, 19]. Concurrently, rapid progress in machine learning and artificial intelligence has enabled the development of algorithms capable of learning complex visual representations directly from medical images. Because pathological diagnosis is fundamentally based on pattern recognition, histopathology has become one of the most promising application areas for artificial intelligence, giving rise to the rapidly expanding field of computational pathology [20-24].

The application of deep learning to colorectal histopathology has produced particularly encouraging results. Modern deep learning architectures are capable of automatically extracting hierarchical image features without requiring handcrafted descriptors, allowing them to identify subtle morphological patterns associated with disease [25, 26]. Consequently, deep learning has been successfully applied to a wide range of tasks, including tumor detection, tissue classification, polyp characterization, molecular subtype prediction, and prognostic assessment [9, 10, 27-29]. Among the most widely investigated approaches, CNNs, especially residual learning architectures [30-32], have consistently demonstrated strong performance across multiple colorectal cancer datasets [9, 10, 27]. More recently, transformer-based models and hybrid CNN-transformer frameworks have further expanded the capabilities of deep learning systems by capturing both local morphological details and broader tissue context [33-36].

As the field has matured, research has evolved beyond simple binary discrimination between benign and malignant tissue. Increasing attention has been directed toward fine-grained histopathological characterization, where models are required to distinguish among multiple lesion types and disease stages [37-39]. Patch-based learning strategies, transfer learning techniques, attention mechanisms, and feature-fusion approaches have all contributed to improved discrimination of subtle morphological differences between colorectal tissue subtypes [40-45]. In parallel, semantic segmentation methods have enabled detailed tissue compartment analysis through pixel-level annotation, providing valuable spatial information for risk stratification and disease assessment [45-47]. Furthermore, recent studies have emphasized the development of lightweight and explainable artificial intelligence frameworks, incorporating techniques such as Gradient-weighted Class Activation Mapping (Grad-CAM) to improve transparency, interpretability, and clinical trust while reducing computational demands [48-53].

Despite these advances, several challenges remain. Many existing studies focus primarily on binary classification tasks or evaluate only a limited number of model architectures, making it difficult to establish which approaches generalize most effectively across diverse histopathological categories. Moreover, although recent datasets have introduced more detailed histological differentiation categories and richer annotations [54, 55], comparatively few studies have systematically investigated the ability of modern deep learning architectures to distinguish among multiple clinically relevant colorectal tissue classes. The relative performance of contemporary transformer-based networks compared with established CNN architectures also remains insufficiently explored in this context. Given the increasing importance of accurate, efficient, and clinically applicable computational pathology tools, comprehensive comparative evaluations are essential.

Therefore, the present study conducts a systematic comparison of twelve state-of-the-art deep learning architectures, encompassing both conventional convolutional neural networks and modern transformer-based models, for colorectal histopathological image classification using the Kather histology dataset [39, 56]. By evaluating these architectures under a consistent experimental framework, this work aims to provide a clearer understanding of their relative strengths, limitations, and potential suitability for future clinical deployment.

# Method

## 2.1. Dataset

This study utilized the publicly available Kather colorectal histology dataset; a benchmark dataset widely used for evaluating deep learning models in computational pathology. The dataset consists of 5,000 hematoxylin and eosin (H&E)-stained histopathological image tiles extracted from colorectal tissue specimens. Images are categorized into eight distinct tissue classes representing both tumor and non-tumor microenvironmental components.

The dataset is balanced, containing 625 image tiles for each tissue category, resulting in a total of 5,000 images. The tissue classes include tumor epithelium, stroma, complex stroma, lymphocyte-rich regions, debris, mucosa, adipose tissue, and empty background regions. A summary of the dataset composition is presented in *Table 1*.

Table 1: Distribution of tissue classes in the Kather colorectal histology dataset.

| Tissue Class | Description | Number of Images |
|---|---|---|
| **Tumor** | Colorectal tumor epithelium | 625 |
| **Stroma** | Connective tissue and stromal regions | 625 |
| **Complex** | Complex stromal structures | 625 |
| **Lympho** | Lymphocyte-rich tissue regions | 625 |
| **Debris** | Necrotic and cellular debris regions | 625 |
| **Mucosa** | Normal colorectal mucosal tissue | 625 |
| **Adipose** | Adipose tissue | 625 |
| **Empty** | Background and empty tissue regions | 625 |
| **Total** | | **5,000** |

The balanced distribution of classes minimizes class imbalance bias during model training and enables a fair comparison among the evaluated deep learning architectures. The dataset encompasses a diverse range of histological patterns commonly encountered in colorectal tissue analysis, making it suitable for benchmarking convolutional neural networks, transformer-based models, and hybrid deep learning architectures. Representative examples of the eight tissue categories included in the dataset are shown in *Figure 1*, illustrating the morphological diversity of colorectal histopathological tissues used for model development and evaluation.

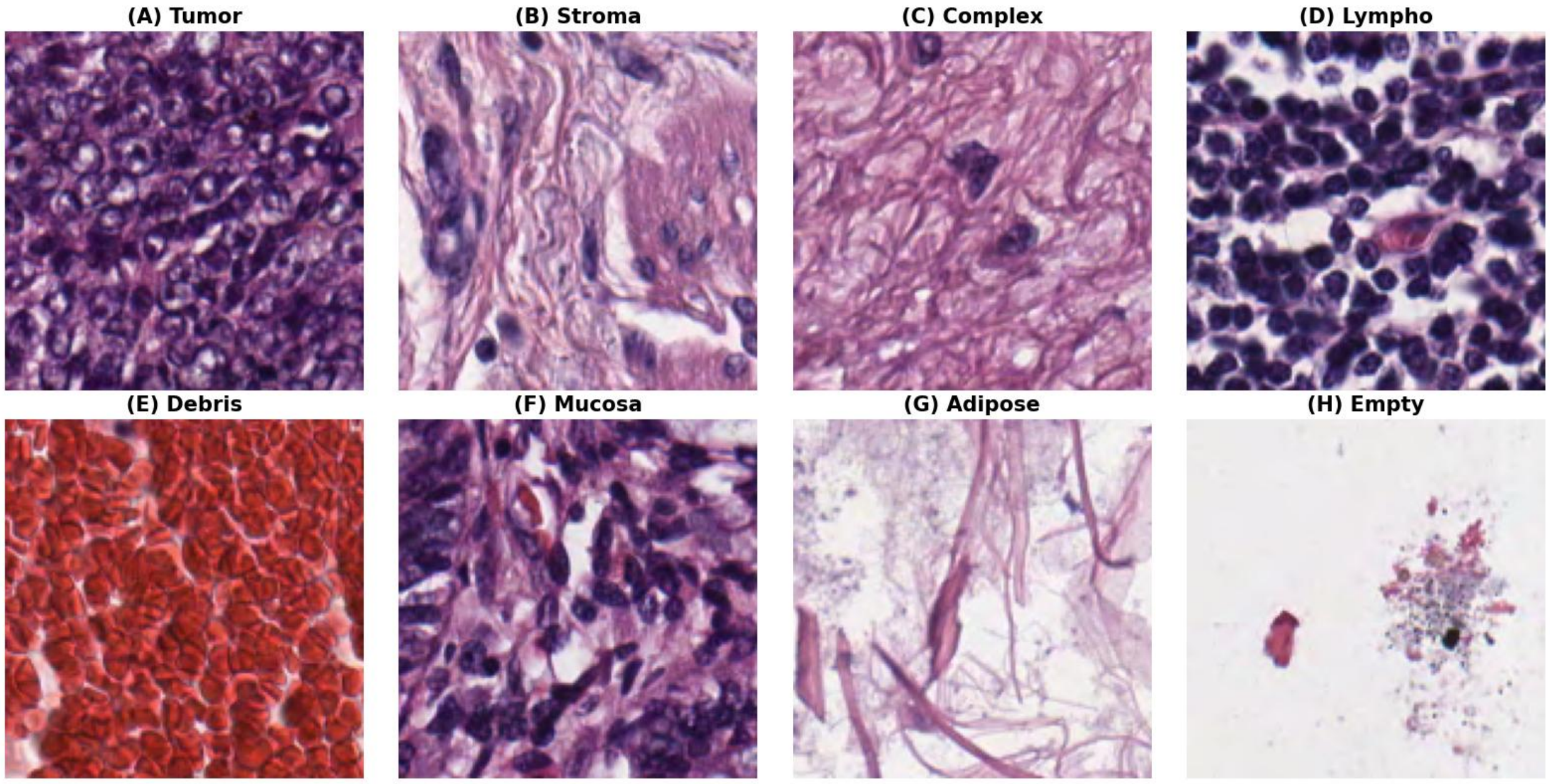

Figure 1: Representative H&E-stained image tiles from the Kather colorectal histology dataset. Example image patches are shown for the eight tissue classes included in this study: (A) Tumor, (B) Stroma, (C) Complex stroma, (D) Lymphocyte-rich tissue, (E) Debris, (F) Mucosa, (G) Adipose tissue, and (H) Empty/background regions.

## 2.2. Data Preprocessing

Prior to model training, all histopathological images underwent a standardized preprocessing pipeline to ensure compatibility with the evaluated deep learning architectures. Images were loaded as three-channel RGB inputs and organized according to their corresponding tissue class labels. Because the evaluated architectures have different input size requirements, two image resolutions were employed. Convolutional neural network (CNN)-based models, including ResNet34, ResNet50, DenseNet121, EfficientNet-B0, and ConvNeXt-Tiny, were trained using images resized to 150 × 150 pixels. Transformer-based and hybrid architectures, including ViT-B/16, Swin Transformer, BEiT, EVA-02, TinyViT, CoAtNet, and NextViT, were trained using images resized to 224 × 224 pixels to match their pretrained model specifications. To improve model generalization and reduce overfitting, online data augmentation was applied during training. Augmentation operations included random horizontal and vertical flipping as well as geometric transformations. These augmentations increased the diversity of training samples while preserving the underlying histopathological characteristics of the tissue images. All models were initialized using ImageNet-pretrained weights. Consequently, image normalization was performed using the standard ImageNet normalization statistics to maintain consistency with the pretraining distribution. The same preprocessing pipeline was applied across all experiments to ensure a fair comparison among the evaluated CNN, transformer-based, and hybrid architectures. A summary of the preprocessing strategy is provided in *Table 2*.

Table 2: Image preprocessing and augmentation pipeline.

| Preprocessing Step | Description |
|---|---|
| **Color Format** | RGB images |
| **CNN Input Size** | 150 × 150 pixels |
| **Transformer/Hybrid Input Size** | 224 × 224 pixels |
| **Data Augmentation** | Random horizontal and vertical flipping; geometric transformations |
| **Weight Initialization** | ImageNet-pretrained weights |
| **Normalization** | ImageNet normalization statistics |

The preprocessing pipeline was designed to preserve diagnostically relevant histological features while providing standardized inputs suitable for deep learning model training and evaluation. The overall study workflow, including dataset preparation, preprocessing, model development, training, evaluation, and interpretability analysis, is illustrated in *Figure 2*.

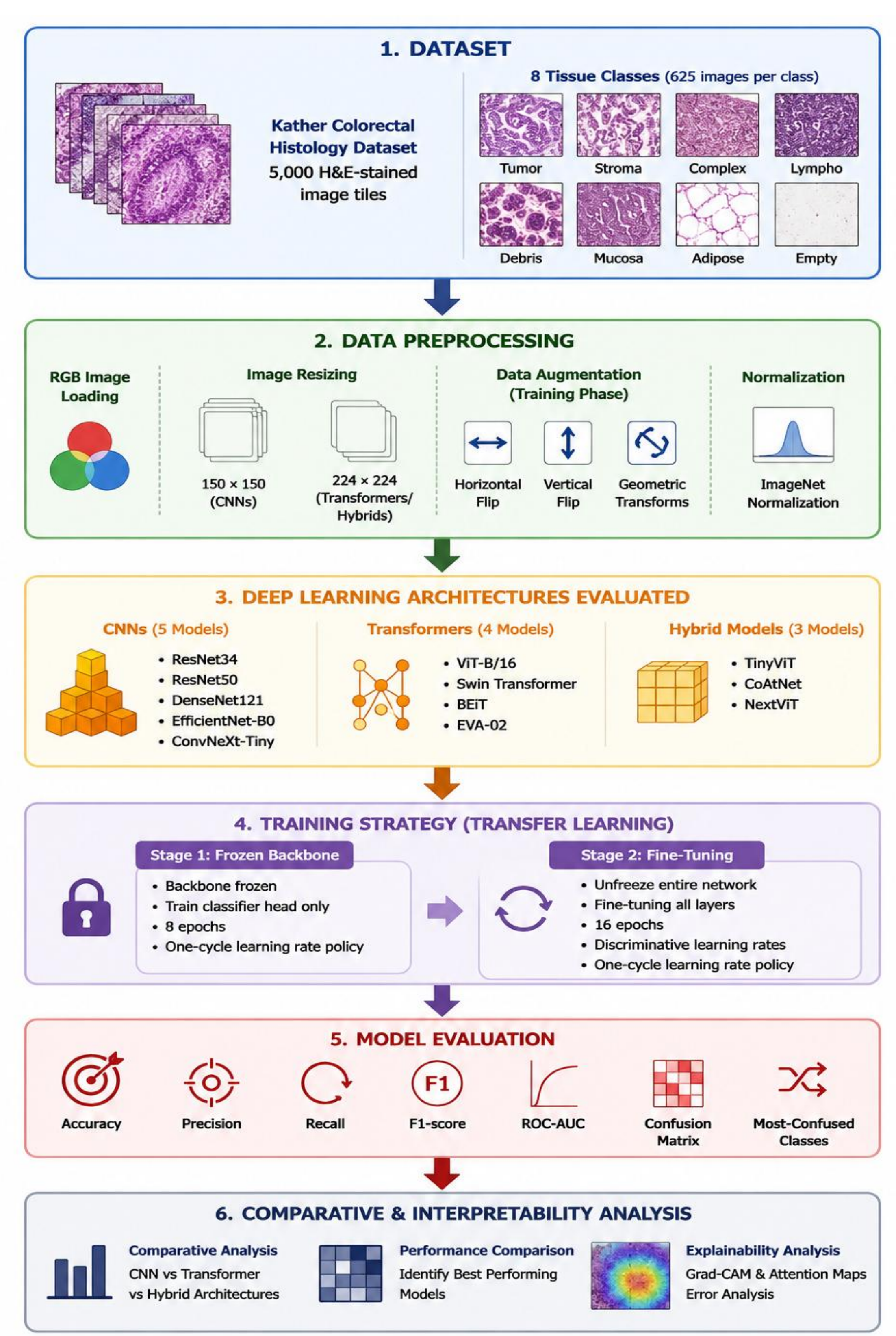

Figure 2: Overview of the study workflow for colorectal histopathology classification, including data preprocessing, model training, evaluation, and interpretability analysis.

## 2.3. Deep Learning Architectures

To comprehensively evaluate the impact of architectural design on colorectal histopathology classification, a diverse set of deep learning models spanning three major categories, CNNs, transformer-based architectures, and hybrid convolution-transformer models, was investigated. All models were initialized using ImageNet-pretrained weights and adapted to perform eight-class tissue classification by replacing the final classification layer with a task-specific output layer. The evaluated architectures are summarized in Table 3.

Table 3: Deep learning architectures evaluated in this study.

| Architecture Family | Model |
|---|---|
| **CNNs** | ResNet34 |
| **CNNs** | ResNet50 |
| **CNNs** | DenseNet121 |
| **CNNs** | EfficientNet-B0 |
| **CNNs** | ConvNeXt-Tiny |
| **Transformer-Based Models** | ViT-B/16 |
| **Transformer-Based Models** | Swin Transformer-Tiny |
| **Transformer-Based Models** | BEiT |
| **Transformer-Based Models** | EVA-02 |
| **Hybrid Architectures** | TinyViT |
| **Hybrid Architectures** | CoAtNet |
| **Hybrid Architectures** | NextViT |

The evaluated architectures were grouped into three major categories: CNNs, transformer-based models, and hybrid architectures, as illustrated in *Figure 3.* This categorization enables a systematic comparison of different deep learning paradigms for colorectal histopathology image classification. CNN-based architectures represent the conventional approach to visual recognition and rely on hierarchical convolutional feature extraction. These models are particularly effective at capturing local spatial patterns and texture information, which are important characteristics of histopathological images. In this study, five CNN architectures were evaluated, ranging from classical residual networks to more recent convolutional designs. Transformer-based architectures utilize self-attention mechanisms to model long-range dependencies and global contextual relationships within images. Unlike CNNs, which primarily focus on local receptive fields, transformers can capture interactions between distant image regions, potentially improving tissue characterization and classification performance. Four transformer-based models were included in the benchmark.

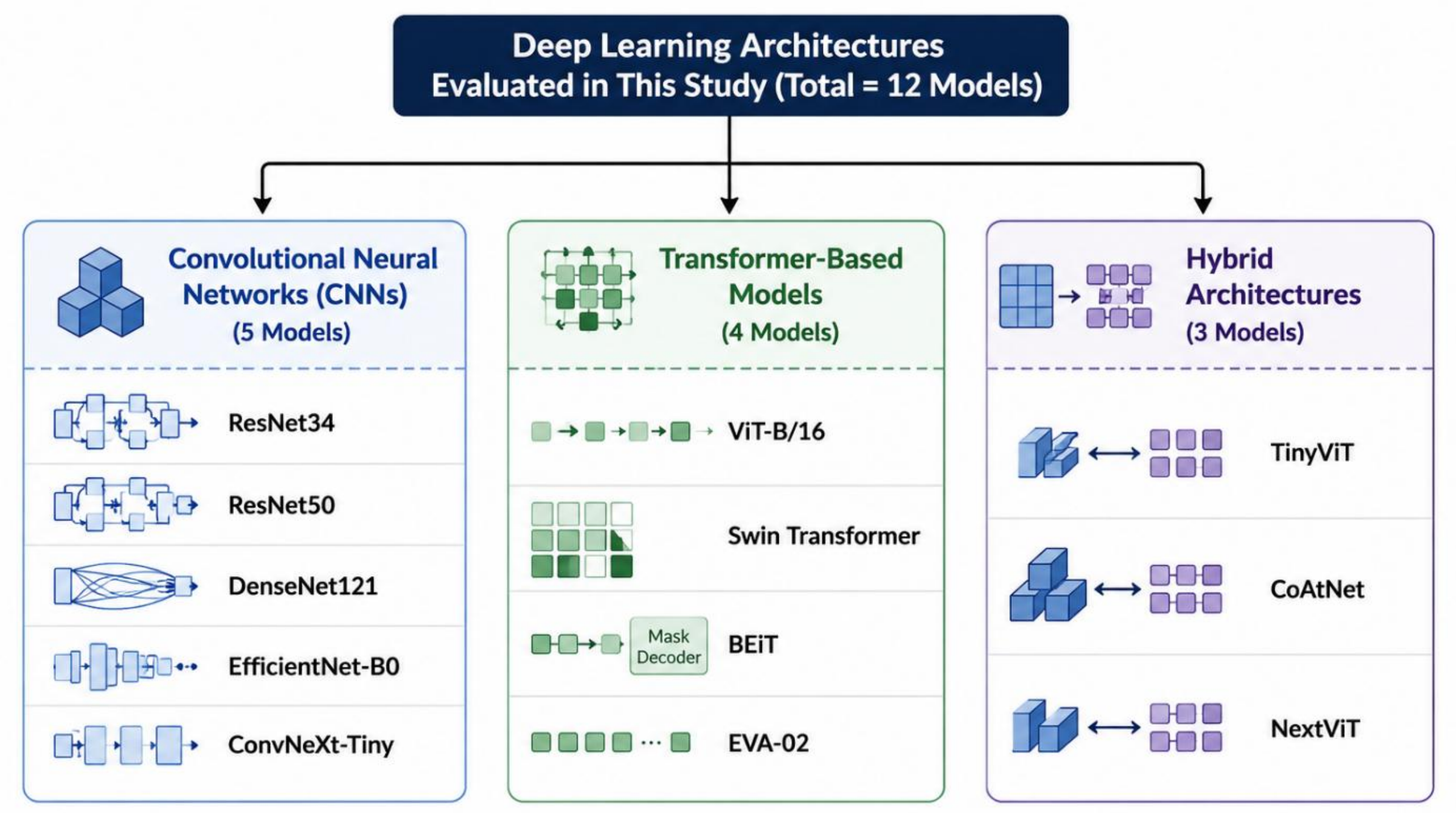

Figure 3: Taxonomy of the deep learning architectures evaluated in this study. The benchmark included twelve models spanning three major families: convolutional neural networks (CNNs), transformer-based models, and hybrid architecture.

Hybrid architectures combine convolutional operations and attention mechanisms to leverage the complementary strengths of both paradigms. These models aim to preserve the strong local feature extraction capability of CNNs while incorporating the global contextual modeling provided by transformers. Three hybrid architectures were evaluated to investigate whether combining these approaches offers advantages for colorectal histopathology classification. By evaluating architectures from these three categories under a unified training and evaluation framework, this study provides a comprehensive comparison of contemporary deep learning paradigms for colorectal histopathology image classification.

## 2.4. Transfer Learning

Transfer learning was employed to leverage knowledge learned from large-scale natural image datasets and adapt pretrained deep learning models to the colorectal histopathology classification task. All evaluated architectures were initialized using weights pretrained on the ImageNet dataset, which contains over one million annotated natural images across one thousand object categories. To adapt each architecture to the target task, the original classification layer was replaced with a task-specific output layer corresponding to the eight tissue classes of the Kather colorectal histology dataset. This modification enabled the pretrained models to perform multiclass histopathological tissue classification while retaining the visual representations learned during pretraining. A uniform transfer learning strategy was applied across all twelve architectures to ensure a fair comparison. Initially, the pretrained feature extraction backbone was frozen, and only the newly added classification head was trained. Subsequently, the entire network was unfrozen and fine-tuned on the colorectal histology dataset using discriminative learning rates. This two-stage approach allowed the models to gradually adapt their learned representations from natural images to histopathological image analysis while reducing the risk of overfitting and catastrophic forgetting. By applying the same transfer learning protocol to all evaluated architectures, performance differences observed in this study can be primarily attributed to architectural characteristics rather than variations in model initialization or training methodology.

### 2.5. Model Training Strategy

All deep learning models were developed using the FastAI framework with PyTorch as the underlying deep learning backend. To ensure a fair comparison among architectures, a standardized training protocol was applied across all experiments. Model training was performed using a two-stage transfer learning strategy. During the first stage, the pretrained backbone of each network was frozen and only the newly added classification head was trained. This stage was conducted for eight epochs using the one-cycle learning rate policy. Optimal learning rates were determined individually for each architecture using the FastAI learning rate finder. Following initial training, all network layers were unfrozen and fine-tuned to enable adaptation of the pretrained feature representations to colorectal histopathology images. Fine-tuning was performed for sixteen additional epochs using discriminative learning rates, allowing lower learning rates for earlier layers and higher learning rates for later layers. The one-cycle learning rate scheduling strategy was maintained throughout the fine-tuning process. To ensure consistency across architecture, all models were trained using the same dataset partitions, preprocessing pipeline, augmentation strategy, and optimization framework. This standardized training methodology enabled a direct comparison of CNN-based, transformer-based, and hybrid architectures while minimizing potential confounding effects arising from differences in training procedures. A summary of the training protocol is provided in *Table 4*.

Table 4: Training strategy employed for all evaluated architectures.

| **Training Stage** | **Configuration** |
|---|---|
| **Framework** | FastAI (PyTorch backend) |
| **Weight Initialization** | ImageNet pretrained weights |
| **Stage 1** | Frozen backbone training |
| **Stage 1 Epochs** | 8 |
| **Stage 2** | Fine-tuning (all layers unfrozen) |
| **Stage 2 Epochs** | 16 |
| **Learning Rate Selection** | Learning rate finder |
| **Learning Rate Policy** | One-cycle learning rate scheduling |
| **Fine-Tuning Strategy** | Discriminative learning rates |

## Results

Figure 4 compares the performance of the twelve deep learning architectures using precision, sensitivity, specificity, ROC-AUC, Cohen's kappa, and MCC. Transformer-based models generally achieved the best results, with EVA-02 and ViT-B/16 consistently ranking highest across most metrics. Among CNNs,

ResNet34 and ConvNeXt-Tiny were the strongest performers, whereas EfficientNet-B0 produced the lowest scores. Hybrid architecture demonstrated intermediate performance. Specificity remained high for all models, and the overall ranking was largely consistent across all metrics, indicating robust and reliable performance differences among the evaluated architectures.

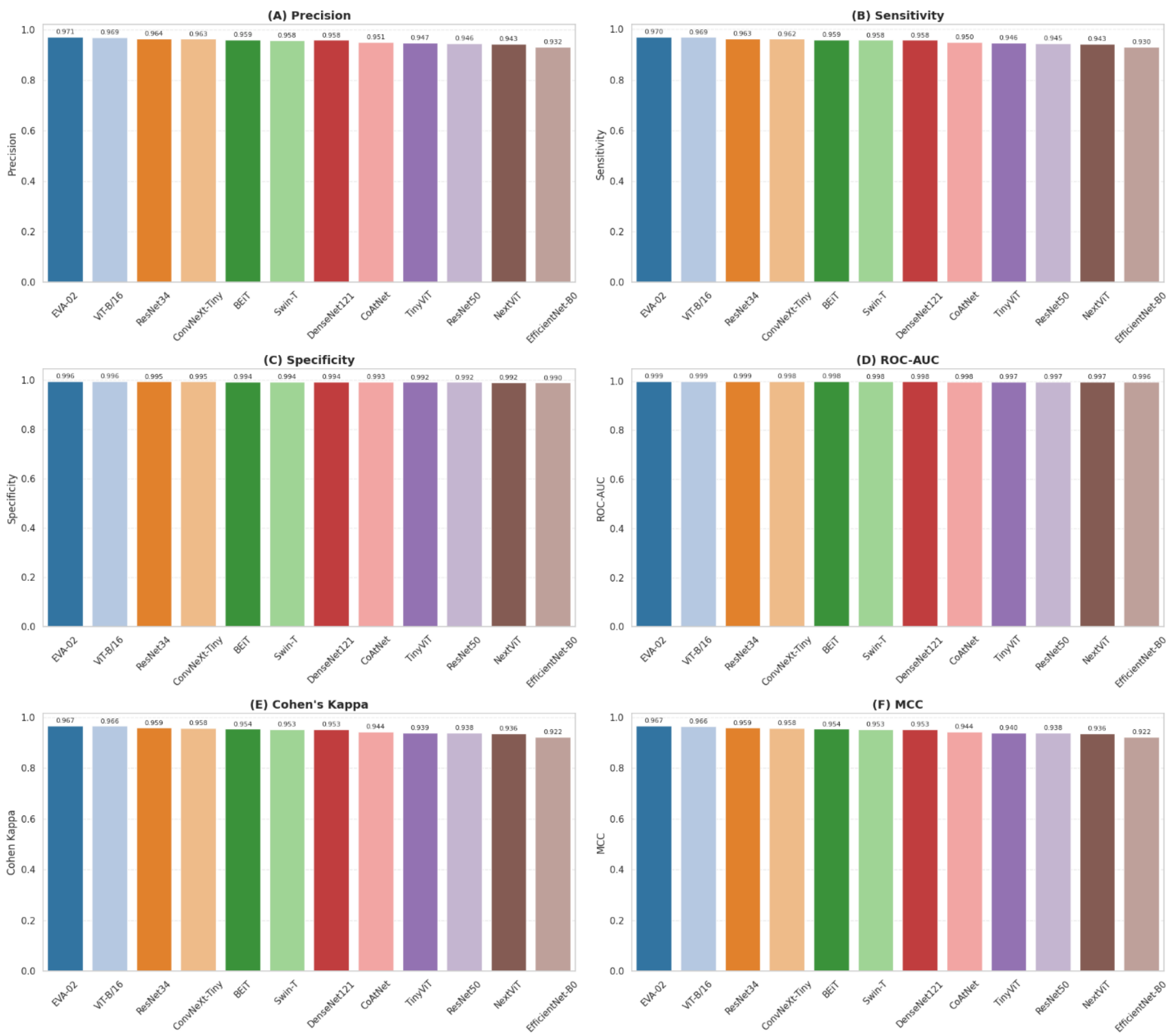

Figure 4: Comparison of evaluation metrics across deep learning architectures

Figure 5 and 6 present the distributions of validation accuracy and F1-score across four training stages (epochs 0-3, 4-7, 8-11, and 12-15). Both metrics increased significantly throughout training ($p < 0.05$ to $p < 0.001$), with the largest gains occurring during the early and middle stages. EVA-02 and ViT-B/16 achieved the highest accuracy and F1-scores, followed by ResNet34 and ConvNeXt-Tiny. DenseNet121 and Swin Transformer-Tiny showed intermediate performance, whereas EfficientNet-B0, NextViT, and TinyViT consistently achieved lower scores. Similar trends across both metrics indicate stable and progressive improvements in classification performance.

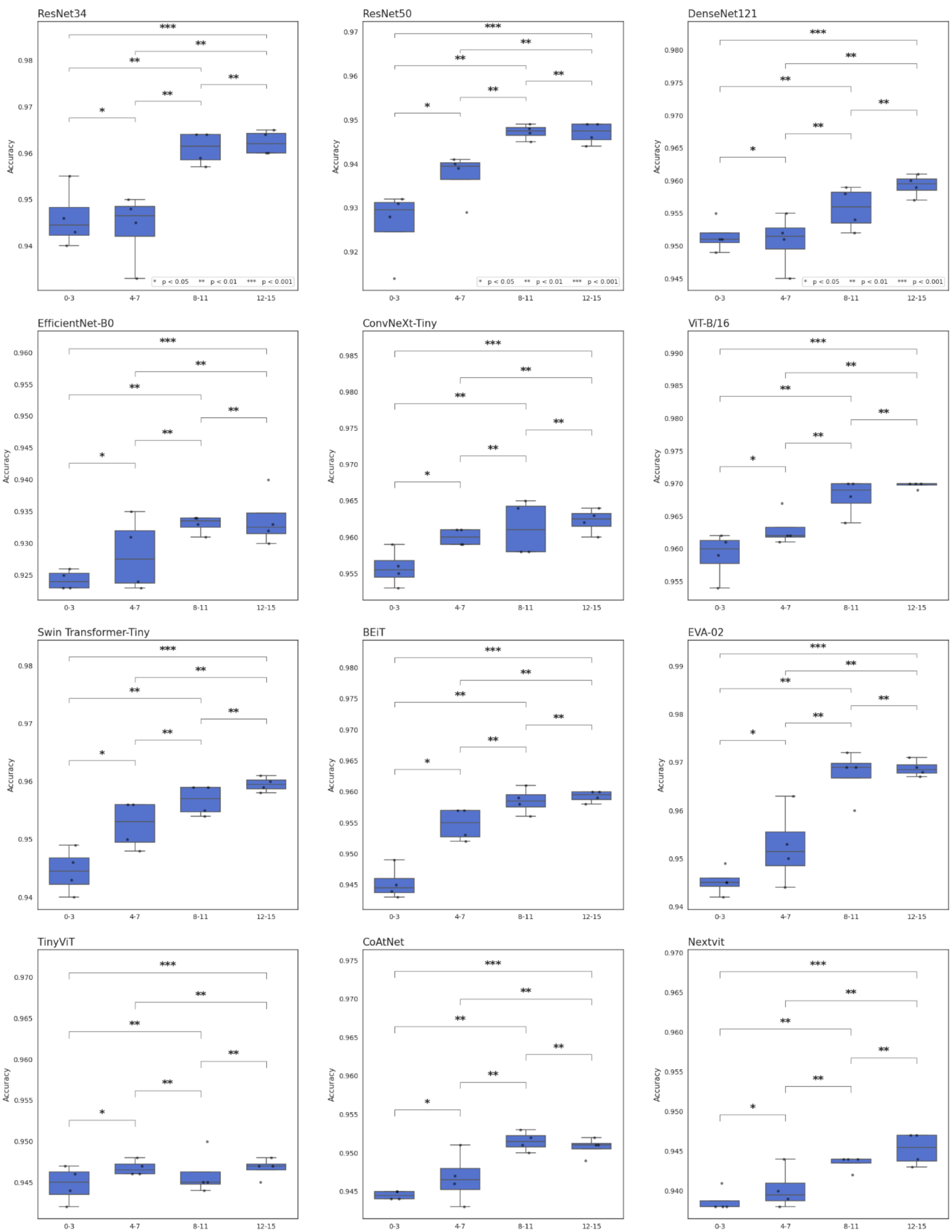

Figure 5: Distribution of validation accuracies across training epochs for the evaluated CNN, transformer-based, and hybrid deep learning architectures on the colorectal histopathology classification task.

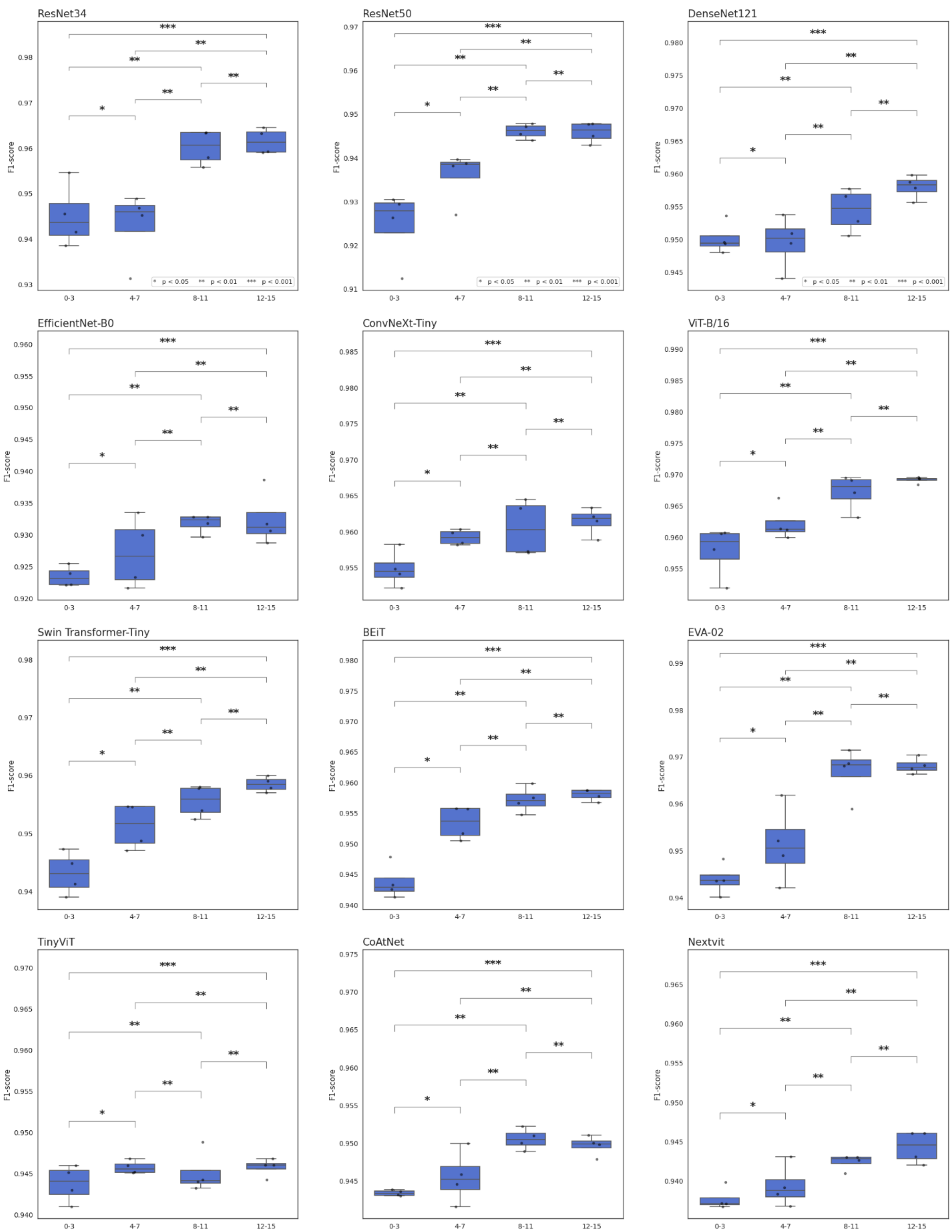

Figure 6: Distribution of validation accuracies across training epochs for the evaluated CNN, transformer-based, and hybrid deep learning architectures on the colorectal histopathology classification task.

To further evaluate the class-level performance of the best-performing architecture, a detailed analysis of the EVA-02 model was conducted. Table 5 summarizes the precision, sensitivity, and F1-score achieved for each histopathological tissue category. This analysis provides additional insight into the model's ability to discriminate between morphologically distinct tissue structures beyond the overall performance metrics presented in Figure 4 to Figure 6.

Table 5: Per-class classification performance of the EVA-02 model on the colorectal histopathology dataset.

| Tissue Class | Precision | Sensitivity (Recall) | F1-score | Support |
|---|---|---|---|---|
| **Tumor** | 0.9520 | 0.9835 | 0.9675 | 121 |
| **Stroma** | 0.9550 | 0.9298 | 0.9422 | 114 |
| **Complex Stroma** | 0.9008 | 0.9147 | 0.9077 | 129 |
| **Lymphocytes** | 0.9593 | 0.9672 | 0.9633 | 122 |
| **Debris** | 0.9848 | 0.9420 | 0.9630 | 138 |
| **Mucosa** | 1.0000 | 1.0000 | 1.0000 | 116 |
| **Adipose Tissue** | 0.9850 | 0.9776 | 0.9813 | 134 |
| **Empty Background** | 0.9767 | 1.0000 | 0.9882 | 126 |

The EVA-02 model achieved consistently high performance across all tissue categories, with F1-scores exceeding 0.90 for every class. Perfect classification performance was obtained for the Mucosa class, while Empty Background and Adipose Tissue also demonstrated near-perfect discrimination. In contrast, Complex Stroma exhibited the lowest precision, sensitivity, and F1-score, indicating that this class represented the most challenging tissue category for classification. Nevertheless, the overall results confirm the strong capability of EVA-02 to accurately identify diverse colorectal histopathological patterns.

To provide a comprehensive comparison of model performance across multiple evaluation criteria, a heatmap was generated using the principal classification metrics. Figure 7 summarizes the performance of all evaluated architectures and enables direct comparison of their overall predictive capabilities.

As shown in Figure 7, EVA-02 achieved the highest overall performance across most evaluation metrics, followed closely by ViT-B/16. Among the convolutional architectures, ResNet34 and ConvNeXt-Tiny produced the strongest results, while EfficientNet-B0 consistently achieved the lowest scores. Transformer-based models generally demonstrated superior performance compared with CNN and hybrid architectures, with particularly high values for accuracy, F1-score, Cohen’s kappa, and MCC. Overall, the heatmap confirms the consistency of model rankings across different evaluation metrics and further highlights the effectiveness of EVA-02 for colorectal histopathology classification.

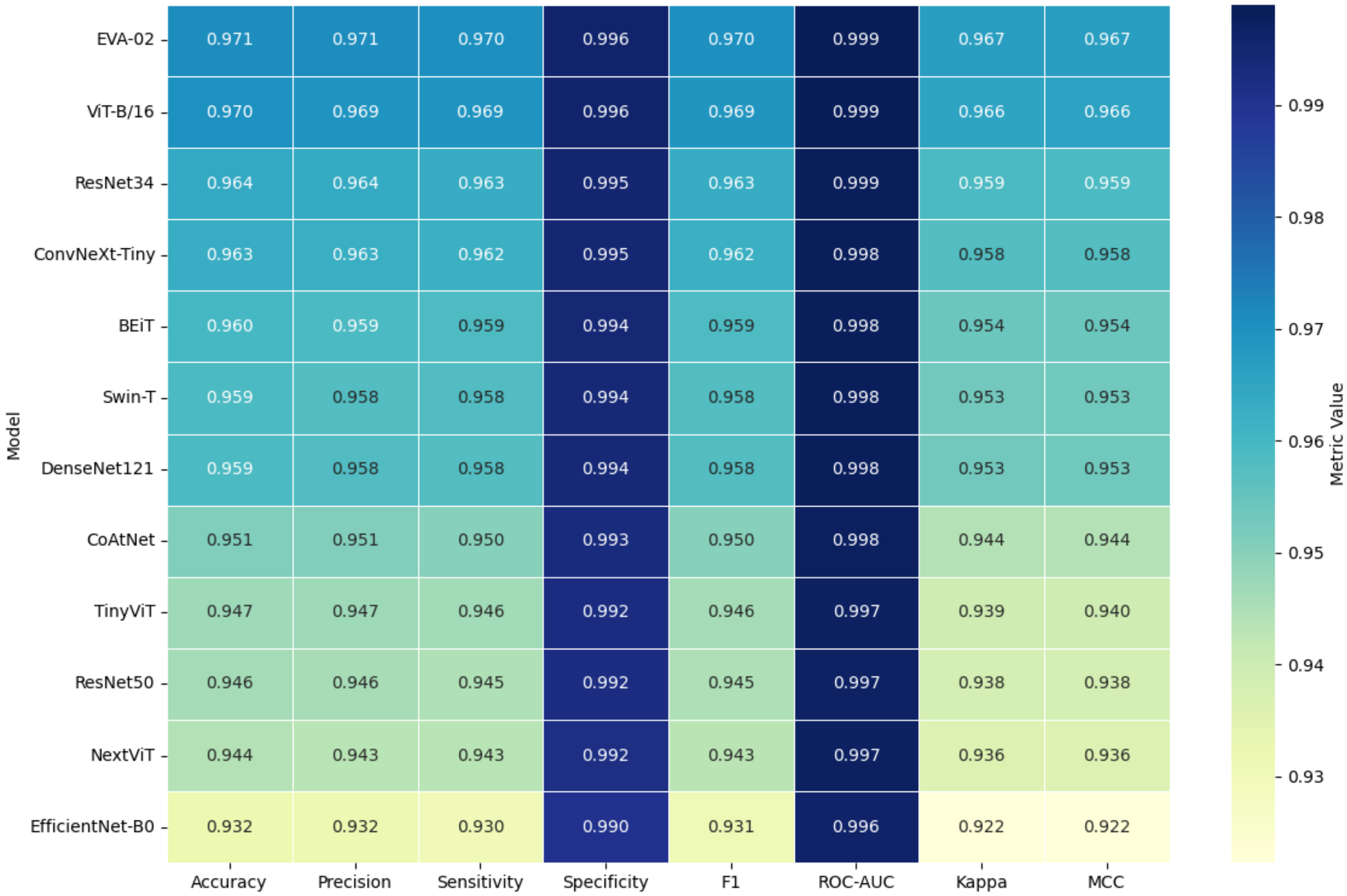

Figure 7: Heatmap comparison of all evaluated deep learning architectures across eight performance metrics, including accuracy, precision, sensitivity, specificity, F1-score, ROC-AUC, Cohen's kappa, and MCC. Higher values indicate better classification performance.

To provide an overall ranking of the evaluated architectures, the two principal performance metrics, accuracy and F1-score, were compared simultaneously. Figure 8 presents a direct comparison of model performance, allowing assessment of both predictive accuracy and classification balance across all evaluated deep learning architectures. As shown in Figure 8, EVA-02 achieved the highest overall performance, obtaining an accuracy of 97.1% and an F1-score of 97.0%, followed closely by ViT-B/16. Among the convolutional architectures, ResNet34 and ConvNeXt-Tiny demonstrated the strongest performance, whereas EfficientNet-B0 achieved the lowest scores. The close agreement between accuracy and F1-score across all models indicates balanced classification behavior and confirms the consistency of the observed ranking. Overall, transformer-based architectures occupied the highest positions, further highlighting their effectiveness for colorectal histopathological image classification.

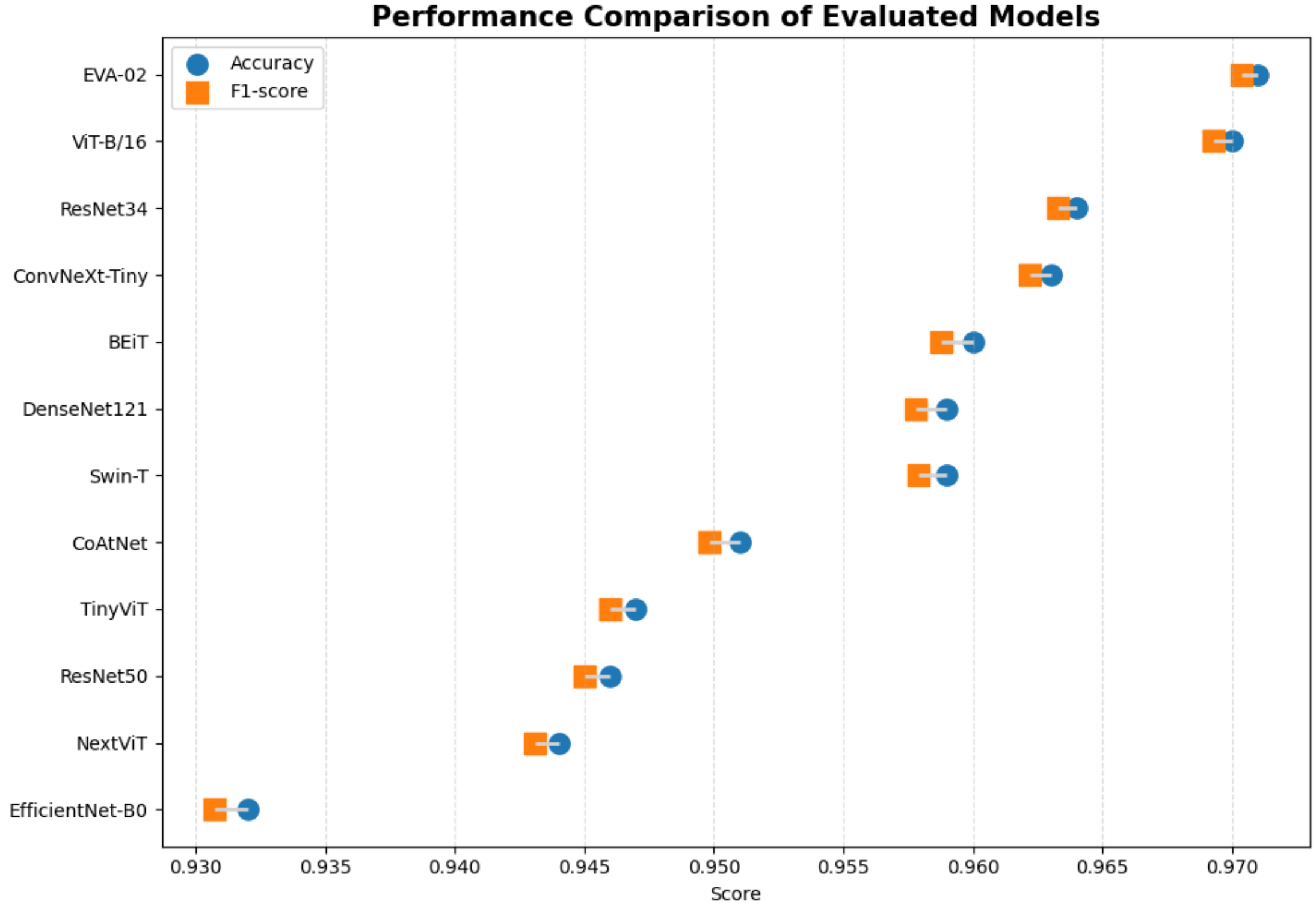

Figure 8: Comparison of the evaluated deep learning architectures based on accuracy and F1-score. Models are ranked according to overall classification performance.

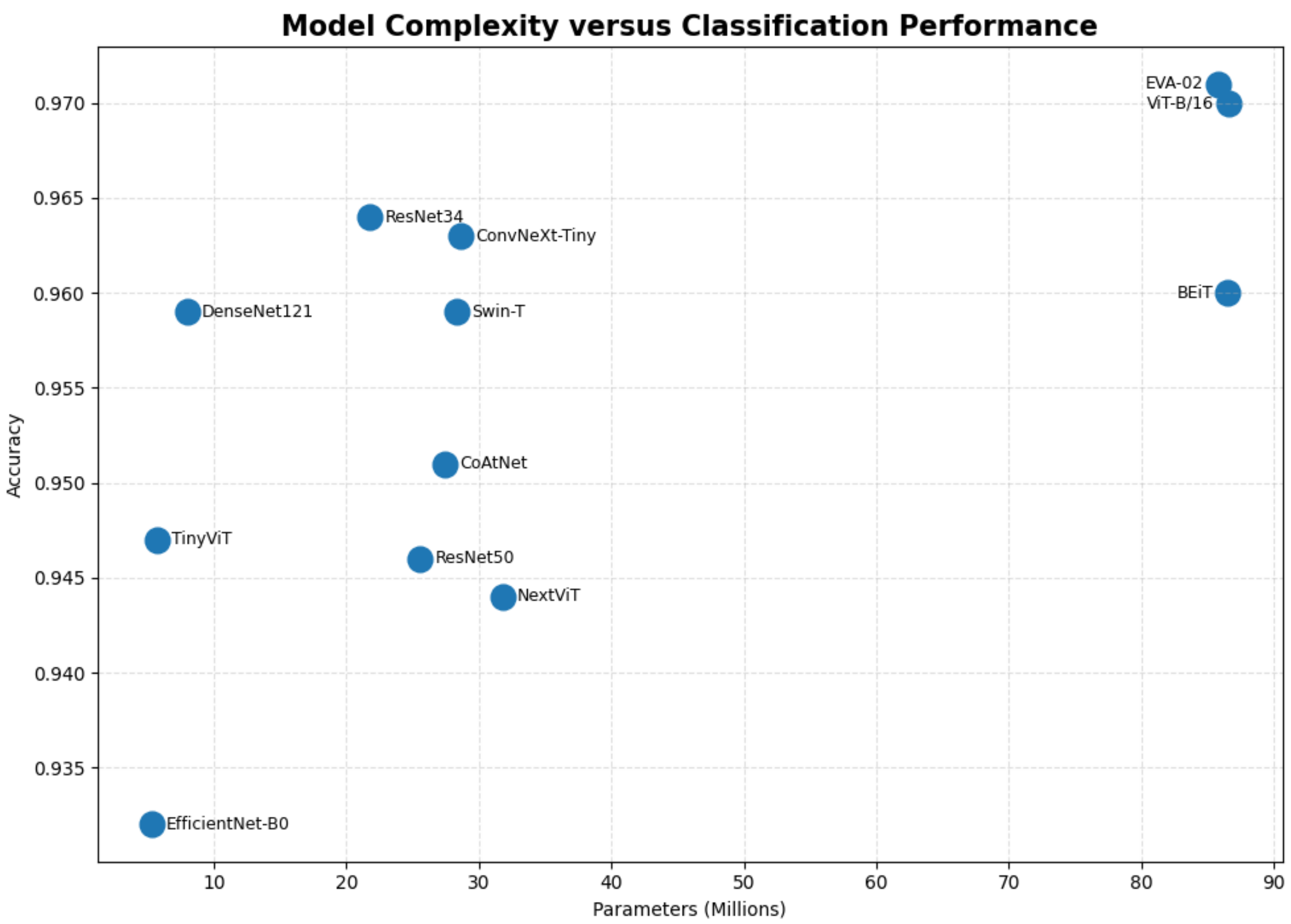

Figure 9: Relationship between model size and classification accuracy for the evaluated architectures.

Figure 9 illustrates the relationship between model complexity and classification performance across the evaluated architectures. EVA-02 and ViT-B/16 achieved the highest accuracies, while ResNet34 and ConvNeXt-Tiny delivered comparable performance with substantially fewer parameters. Overall, these findings suggest that increased model complexity does not necessarily translate into superior classification performance.

## Discussion

Figure 5 and 6 summarize the evolution of validation accuracy and F1-score, respectively, across four training stages for all evaluated architectures. Overall, both figures demonstrate a consistent improvement in predictive performance as training progressed, with statistically significant differences observed between early and later epochs for nearly all models. The close agreement between the trends observed in Figure 5 and 6 indicates that improvements in classification accuracy were accompanied by corresponding gains in the balance between precision and recall, suggesting that the models improved their overall discrimination capability rather than simply favoring dominant classes. Among the CNN-based architectures, ResNet34 achieved the strongest overall performance, reaching a final accuracy of 96.4% and an F1-score of 96.3%. ConvNeXt-Tiny closely followed with an accuracy of 96.3% and an F1-score of 96.2%, while exhibiting stable convergence behavior throughout training. DenseNet121 achieved intermediate performance (95.9% accuracy and 95.8% F1-score), outperforming ResNet50 (94.6% accuracy, 94.5% F1-score) and EfficientNet-B0 (93.2% accuracy, 93.1% F1-score). The relatively poor performance of EfficientNet-B0 compared with the remaining CNNs suggests that lightweight scaling strategies may be less effective than residual and modernized convolutional designs for colorectal histopathology classification. Overall, ResNet34 and ConvNeXt-Tiny emerged as the most competitive convolutional architectures, demonstrating that carefully optimized CNNs remain highly effective for tissue image analysis. Within the transformer-based models, EVA-02 achieved the highest overall performance, reaching 97.1% accuracy and 97.0% F1-score. ViT-B/16 produced nearly identical results, achieving 97.0% accuracy and 96.9% F1-score, and remained consistently among the strongest performers throughout training. Swin Transformer-Tiny and BEiT achieved moderate performance, reaching 95.9% and 95.2% accuracy, respectively. The superior results obtained by EVA-02 and ViT-B/16 suggest that large-scale pretrained transformer representations are highly effective in capturing discriminative histopathological features. Interestingly, despite employing more advanced pretraining strategies, EVA-02 provided only a marginal improvement over the standard Vision Transformer, indicating that the colorectal histopathology dataset may already be well represented by conventional transformer architectures. Among the hybrid convolution-transformer models, CoAtNet achieved the best overall performance, reaching 95.1% accuracy and 95.0% F1-score. TinyViT achieved 94.7% accuracy and 94.6% F1-score, while NextViT reached 94.4% accuracy and 94.3% F1-score. Although these hybrid architectures consistently demonstrated stable improvements during training, none surpassed the strongest CNN or transformer models. Nevertheless, the relatively small performance gap between hybrid and transformer architectures indicates that combining convolutional inductive biases with self-attention mechanisms remains a promising strategy for computational pathology applications. Comparison across architectural families revealed a clear performance hierarchy. Transformer-based models achieved the highest overall predictive performance, with EVA-02 and ViT-B/16 occupying the top two positions in both Figures 4 and 5. The best CNN model, ResNet34, remained highly competitive, trailing EVA-02 by only 0.7 percentage points in accuracy and 0.7 percentage points in F1-score. Similarly, ConvNeXt-Tiny achieved performance comparable to several transformer architectures while maintaining substantially lower computational complexity. These findings indicate that although transformer models provide the strongest predictive performance, modern convolutional architecture continues to offer an attractive balance between accuracy, robustness, and computational efficiency. Importantly, the similarity between the accuracy distributions shown in Figure 5 and the F1-score distributions shown in Figure 6 confirm that the observed performance improvements were not driven by biased predictions toward specific tissue classes. This observation is further supported by the final evaluation metrics, where the leading models exhibited highly consistent values across accuracy, precision, sensitivity, and F1-score. EVA-02 achieved 97.1% accuracy, 97.1% precision, 97.0% sensitivity, and 97.0% F1-score, while ViT-B/16 achieved 97.0%, 96.9%, 96.9%, and 96.9%, respectively. Such consistency indicates strong class-balanced performance and reliable generalization across all colorectal tissue categories.

Additional analysis further supported the superiority of the transformer-based architectures. As shown in Figure 7, EVA-02 consistently achieved the highest scores across all evaluation metrics, including precision, sensitivity, specificity, ROC-AUC, Cohen's kappa, and MCC, confirming that its performance advantage was not limited to a single metric. The high ROC-AUC values observed for all models indicate strong discriminative capability, whereas the superior kappa and MCC scores obtained by EVA-02 and ViT-B/16 demonstrate improved agreement and classification reliability across tissue classes. The per-class analysis of EVA-02 (Table *5*) revealed consistently strong performance across all histopathological categories, with F1-scores exceeding 0.90 for every class. Perfect classification was achieved for the Mucosa class, while Complex Stroma represented the most challenging category, likely reflecting greater morphological similarity to neighboring tissue structures. Furthermore, the ranking analysis presented in Figure 8 confirmed the overall ordering of model performance, with EVA-02 and ViT-B/16 occupying the top positions and EfficientNet-B0 consistently demonstrating the weakest results. Interestingly, the complexity-performance relationship illustrated in Figure 9 showed that increased model size did not necessarily translate into proportional gains in classification accuracy. Although EVA-02 and ViT-B/16 achieved the highest predictive performance, smaller architectures such as ResNet34 and ConvNeXt-Tiny produced highly competitive results with substantially fewer parameters, highlighting the importance of considering computational efficiency alongside predictive performance when selecting models for practical computational pathology applications.

## Conclusion

This study presented a comprehensive benchmark of twelve deep learning architectures representing convolutional neural networks, transformer-based models, and hybrid CNN-transformer frameworks for multiclass colorectal histopathology image classification. All evaluated models achieved strong performance, demonstrating the effectiveness of transfer learning for computational pathology applications. Among the investigated architectures, transformer-based models produced the best overall results, with EVA-02 and ViT-B/16 consistently outperforming competing approaches across multiple evaluation metrics. Nevertheless, modern CNN architectures, particularly ResNet34 and ConvNeXt-Tiny, achieved highly competitive performance while requiring substantially lower model complexity, highlighting their continued practical relevance.

The findings indicate that transformer architectures offer superior feature representation capabilities for colorectal tissue classification, whereas carefully optimized CNN models provide an attractive balance between predictive performance and computational efficiency. These results provide a comprehensive reference for model selection in colorectal histopathology analysis and demonstrate that architectural choice should consider both classification performance and resource requirements. Future studies should investigate external validation on larger multi-center datasets, whole-slide image analysis, and explainable artificial intelligence approaches to further enhance the clinical applicability of deep learning systems in digital pathology.

**Acknowledgments**

The author gratefully acknowledges Jakob Nikolas Kather, Frank Gerrit Zöllner, Francesco Bianconi, Susanne M. Melchers, Lothar R. Schad, Timo Gaiser, Alexander Marx, and Cleo-Aron Weis for developing and publicly releasing the colorectal histopathology dataset used in this study.

**Data Availability**

The data used in this study are publicly available from the colorectal histopathology dataset developed by Kather et al. and cited in the manuscript. All data analyzed during this study can be obtained from the original public repository.

**Conflict of Interest**

The author declares no conflict of interest.

**Funding**

This research received no external funding.